\documentstyle[aps,prb, tighten,preprint,amsmath,graphicx]{revtex}
\begin{document}
\draft \title{Two-probe theory of scanning tunneling microscopy of single molecules: Zn(II)-etioporphyrin on alumina}
\author{John Buker and George Kirczenow}
\address{Department of Physics, Simon Fraser University, Burnaby,
British Columbia, Canada, V5A 1S6}
\date{\today}
\maketitle
\begin{abstract}

We explore theoretically the scanning tunneling microscopy of single
molecules on substrates using a framework of two $local$ probes. This
framework is appropriate for studying electron flow in
tip/molecule/substrate systems where a thin insulating layer between the
molecule and a conducting substrate transmits electrons non-uniformly and
thus confines electron transmission between the molecule and substrate
laterally to a  nanoscale region significantly smaller in size than the
molecule. The tip-molecule coupling and molecule-substrate coupling are
treated on the same footing, as local probes to the molecule, with
electron flow modelled using the Lippmann-Schwinger Green function
scattering technique. STM images are simulated for various positions of
the stationary (substrate) probe below a Zn(II)-etioporphyrin I molecule.
We find that these images have a strong dependence on the substrate probe
position, indicating that electron flow can depend strongly on both tip
position and the location of the dominant molecule-substrate coupling.
Differences in the STM images are explained in terms of the molecular
orbitals that mediate electron flow in each case. Recent experimental
results, showing STM topographs of Zn(II)-etioporphyrin I on
alumina/NiAl(110) to be strongly dependent on which individual molecule
on the substrate is being probed, are explained using this model. 
A further experimental test of the model is also proposed.

\end{abstract}
%\begin{multicols}{2}
\section{Introduction}
\label{Introduction}

In the last 20 years, scanning tunneling microscopy has become an
increasingly valuable tool for studying electron transport
through individual molecules.
Experiments in this area of research
involve the adsorption of molecules onto a substrate
and analysis using an STM tip to probe the system.
Some early examples are found in
Refs. \onlinecite{Ohtani,Lippel,Eigler,Joachim,Datta}.
The last 5-10 years has seen the emergence of a true wealth of such
experiments on many different molecular
systems\cite{Gim99,Ho02}. In a simplified
picture of transport through an STM tip/molecule/substrate
system, when a finite potential bias is applied between the tip and
the substrate, the tip and substrate electrochemical potentials
separate and molecular orbitals located in the window of energy
between the two electrochemical potentials mediate electron
flow between the tip and the substrate. In this way, experimental
data such as current-voltage chracteristics may give clues to the
electronic structure of the molecule. 

A more complete understanding of these tip/molecule/substrate
systems is complicated by the fact that the molecule may interact
strongly with the tip and substrate, and in that case one should not think
of the molecule as being an isolated part of the system. 
Electron flow has been
shown to be dependent on the details of the tip and substrate
molecular
coupling\cite{Fischer,Kenkre95b,Sautet96,Sautet97,Lu,Jung,Schunack}.
In experiments on planar or near-planar
molecules, constant current
or constant height topographic STM images of sub-molecular
resolution\cite{Lippel,Lu,Jung,Walzer,Qiu03,Qiu04a,Dong,Repp05}
show how current flow through a molecule is dependent on the lateral
position of the STM tip above the molecule. These topographic
images may also depend on details of the molecule/substrate
configuration. For example, STM experiments on ``lander" molecules
on a Cu(111) surface\cite{Schunack}
show differences in topographic images when
a molecule is moved towards a step edge. 

Theoretical approaches to modeling STM-based electron flow commonly
treat the tip as a probe of the molecule-substrate system. 
The Bardeen\cite{Bardeen}
approximation considers the tip and sample to be two
distinct systems that are perturbed by an interaction
Hamiltonian. Techniques such as the Tersoff-Hamann
formalism\cite{Tersoff83} calculate
a tunneling current based on the local density of states (LDOS) of
the tip and of the sample. Such approaches are widely used and
have been very productive for the understanding of these systems.
There has correspondingly been much
interest in studying the tip-molecule interaction and the details
of the coupling. Theoretical and experimental results in this area
can be readily compared by comparing real and simulated STM topograph
maps. Details of the molecule-substrate coupling may also affect the
the image of the molecule. A number of different theoretical
approaches\cite{Sautet88,Doyen90,Sacks91,Tsukada91,Vigneron92,Kenkre95}
have been developed that predict effects of molecule-substrate
coupling\cite{Fischer,Kenkre95b,Sautet96,Sautet97,Schunack},
in experimental situations where the geometry of the substrate is
homogeneous.
In many of these experimental situations, a molecule is placed
on a metal substrate, resulting in strong coupling along the
entire molecule-metal interface.
More recently, experimental systems of molecules placed on
thin insulating layers above the metal part of the
substrate have allowed the mapping of HOMO-LUMO orbitals
of the molecule as well as the study of molecular
electroluminescence\cite{Qiu03,Qiu04a,Dong,Repp05,Buker02}.
Some of these systems involve relatively simple substrates,
including an insulating layer that behaves qualitatively like a
uniform tunnel barrier\cite{Dong,Repp05} and considerable progress
has been made understanding their STM images. Others, with planar
molecules on  alumina/metal substrates, have more complex
images\cite{Qiu03,Qiu04a} that depend on the precise location
of the molecule on the substrate and are much less well-understood.
STM images of thin ($5$\AA) $pristine$ alumina films on NiAl(111)
surfaces exhibit regular arrays of bright spots\cite{NiAL111}
that signal locations where the film is the most conductive.  The most
conductive locations are spaced 15 to 45\AA\ apart depending on the bias
voltage applied between the STM tip and substrate.\cite{NiAL111} Thin
alumina films on NiAl(110) surfaces have similar small, relatively
conductive regions, although in this case they do not form simple
periodic patterns, presumably because the structure of the alumina film
is not commensurate with the NiAl(110) substrate.\cite{NiAL110}   Thus it
is reasonable to suppose that for such systems the alumina film behaves
as an {\em non-uniform} tunnel barrier between a molecule on its surface
and a metal substrate beneath it and that electrons are transmitted
between the molecule and substrate primarily at the more conductive spots
of the alumina film. If the adsorbed planar molecule is similar in size
to the average spacing between the most conductive spots of the alumina
film (this is the case for the Zn(II)-etioporphyrin I molecules studied
experimentally in Ref.\onlinecite{Qiu03}), then a {\em single} conductive
spot of the film can dominate the electronic coupling between a
suitably placed molecule and the underlying metal substrate.  Thus, as is
shown schematically in Fig. \ref{fig1}, in an STM experiment on such a
system not only the STM tip but also the substrate should be regarded as a
$highly$ $local$ probe making direct electrical contact with a small part
of the molecule. Therefore conventional STM experiments on such systems
can in principle yield information similar to that from experiments
probing a single molecule simultaneously with $two$ separate atomic
STM tips, which are beyond the reach of present day technology.    In this
article, we propose a simple approach for modeling such systems that
should be broadly applicable, and use it to explain the results of recent
experiments.\cite{Qiu03} 

We re-examine scanning tunneling microscopy of
molecules, treating the tip-molecule coupling and the
molecule-substrate coupling on the same footing,
both as {\it local} probes of
the molecule, as is shown schematically in Fig. \ref{fig2}. In this
two-probe model, the probes are represented using a one-dimensional
tight-binding model, and electron flow is modelled using the
Lippmann-Schwinger Green function scattering technique. We find that
the STM image of a molecule can be sensitive to the location of the
dominant molecule-substrate coupling.

We present results for the Zn(II)-etioporphyrin I molecule, treated with
extended H\"{u}ckel theory. STM-like images
are created by simulating movement of the tip probe laterally above
the molecule while keeping the substrate probe at a fixed position
below the molecule. We obtain different current maps for various
positions of the stationary (substrate) probe, and explain their
differences in terms of the molecular orbitals that mediate
electron flow in each case. Our results are shown to be consistent
with recent experimental STM imagery for the system
of Zn(II)-etioporphyrin I on an alumina-covered NiAl(110)
substrate\cite{Qiu03}. By using the two-probe approach described in
this article, we are able to account for all of
the differing types of topographic maps that are seen when this
molecule is adsorbed at different locations on the substrate.
However, despite the success of our model in accounting for the
observed behavior of this system, we emphasize that a detailed microscopic
knowledge of exactly how the Zn(II)-etioporphyrin I molecules interact
with the alumina-covered NiAl surface is still lacking and we hope that
the present study will stimulate further experimental/theoretical
elucidation of this system. We propose an experiment that may shed
additional light on this issue at the end of this article.

\section{The Model}

In the present model,
the tip and substrate are represented by probes, with each probe modelled
as a one-dimensional tight-binding chain, as is depicted in Fig.
\ref{fig3}. The molecule is positioned between the probes, so that it
mediates electron flow between the tip and substrate. The model
Hamiltonian of this system can be divided into three parts, 
$H=H_{probes}+H_{molecule}+W$, where $W$ is the interaction Hamiltonian
between the probes and the molecule. The Hamiltonian for the probes is
given by
\begin{align}
H_{probes} &= \sum_{n=-\infty}^{-1}\epsilon_{tip}|n\rangle\langle
n|+\beta(|n\rangle\langle n-1|+|n-1\rangle\langle n|)\nonumber\\
&+\sum_{n=1}^{\infty}\epsilon_{substrate}|n\rangle\langle n|+\beta 
(|n\rangle\langle n+1|+|n+1\rangle\langle n|),
\label{Hprobes}
\end{align}
where $\epsilon_{tip}$ and $\epsilon_{substrate}$ are
the site energies of the tip and substrate probes,
$\beta$ is the hopping amplitude between nearest neighbour
probe atoms, and $|n\rangle$ represents the orbital at site
$n$ of one of the probes. We take the electrochemical
potentials of the tip and substrate probes to be
$\mu_T=E_F+eV_{bias}/2$ and $\mu_S=E_F-eV_{bias}/2$, where
$V_{bias}$ is the bias voltage applied between them and $E_F$
is their common Fermi level at zero applied bias. The applied
bias also affects the site energies $\epsilon_{tip}$
and $\epsilon_{substrate}$ so that 
$\epsilon_{tip}=\epsilon_{0,tip}+eV_{bias}/2$ and
$\epsilon_{substrate}=\epsilon_{0,substrate}-eV_{bias}/2$,
where $\epsilon_{0,tip}$ and $\epsilon_{0,substrate}$
are the site energies of the tip and substrate probes at
zero bias. In this model, the potential drop from the tip probe
to the molecule, and from the molecule to the substrate, are
assumed to be equal, and there is no potential drop within
the molecule.\cite{drop} Thus, the molecular orbital energies
are considered to be fixed when a bias voltage is applied. 
The Hamiltonian of the molecule
may be expressed as 
\begin{equation}
H_{molecule}= \sum_{j}\epsilon_j|\phi_j\rangle\langle\phi_j|,
\label{Hmol}
\end{equation}
where $\epsilon_j$ is the energy of the $j^{th}$ molecular orbital
($|\phi_j\rangle$). The interaction Hamiltonian between the probes
and molecule is given by
\begin{equation}
W = \sum_{j}W_{-1,j}|-1\rangle\langle\phi_j|
+ W_{j,-1}|\phi_j\rangle\langle -1|
+W_{j,1}|\phi_j\rangle\langle 1|
+W_{1,j}|1\rangle\langle\phi_j|,
\label{Hint}
\end{equation}
where $W_{-1,j}$, $W_{j,-1}$, $W_{j,1}$ and $W_{1,j}$ are the
hopping amplitude matrix elements between the probes and the
various molecular orbitals $|\phi_j\rangle$. 

Electrons initially propagate through one of the probes (which we will
assume to be the tip probe) toward the molecule in the form of
Bloch waves, and may
either undergo reflection or transmission when they encounter the molecule.
Their wavefunctions are of the form
\begin{equation}
|\psi\rangle=\sum_{n=-\infty}^{-1}(e^{iknd} +
re^{-iknd})|n\rangle+\sum_{n=1}^{\infty}te^{ik^\prime
nd}|n\rangle+\sum_{j}c_{j}|\phi_{j}\rangle
\label{psi}
\end{equation}
where $d$ is the lattice spacing, and $t$ and $r$ are the transmission
and reflection coefficients. Upon transmission, the
wavevector $k$ changes to $k^{\prime}$ due to the difference in
site energies $\epsilon_{Tip}$ and $\epsilon_{Substrate}$ of the
tip and substrate probes. 
The transmission probability is given by  
\begin{equation}
T=|t|^2\left|\frac{v(k^\prime d)}{v(kd)}\right|=|t|^2\frac{sin(k^\prime d)}{sin(kd)}
\label{transmit}
\end{equation}
where $v(k)$ and $v(k^{\prime})$ are the respective velocities
of the incoming and transmitted waves.

The transmission amplitude $t$ may be evaluated by solving
a Lippmann-Schwinger equation for this system,
\begin{equation}
|\psi\rangle=|\phi_{0}\rangle+G_{0}(E)W|\psi\rangle,
\label{lippmann}
\end{equation}
where $G_0(E)=(E-(H_{probes}+H_{molecule})+i\delta)^{-1}$
is the Green function for the decoupled system
(without $W$), and $|\phi_0\rangle$ is the eigenstate of an 
electron in the decoupled tip probe. $G_0(E)$ may be separated into 
the three decoupled components: the tip and substrate probes, and the 
molecule. For the tip/substrate probes,
\begin{equation}
G_0^{Tip/Substrate} = \sum_k\frac{|\phi_0(k)\rangle\langle\phi_0(k)|}
{E-(\epsilon_{Tip/Substrate}+2\beta cos(kd))}
\label{gprobe}
\end{equation}
where $d$ is the lattice spacing and 
$\epsilon_{Tip/Substrate}+2\beta cos(kd)$ is the energy of
a tip/substrate electron with wavevector $k$. For the molecule,
\begin{equation}
G_0^M=\sum_j\frac{|\phi_j\rangle\langle\phi_j|}{E-\epsilon_j}
=\sum_j(G_0^M)_j|\phi_j\rangle\langle\phi_j|.
\label{gmolecule}
\end{equation}
The transmission probability for such a system using this formalism has 
been previously solved\cite{Emberly98}, and found to be equal to
\begin{equation}
T(E) = |\frac{A(\phi_0)_{-1})}{[(1-B)(1-C)-AD]}|^2 \frac{sin(k_0^\prime d)}{sin(k_0d)}
\label{transmit1}
\end{equation} 
where $(\phi_0)_{-1}=\langle -1|\phi_0\rangle$, and
\begin{align}
&A=(e^{ik_0^\prime d}/\beta)\sum_j W_{1,j}(G_0^M)_j W_{j,-1} \nonumber \\
&B=(e^{ik_0^\prime d}/\beta)\sum_j (W_{1,j})^2(G_0^M)_j \nonumber \\
&C=(e^{ik_0d}/\beta)\sum_j (W_{-1,j})^2(G_0^M)_j \nonumber \\
&D=(e^{ik_0d}/\beta)\sum_j W_{-1,j}(G_0^M)_j W_{j,1}.
\label{ABCD}
\end{align}
Here, $k_0$ is the wavevector of an electron in the tip probe
with energy $E$, and $k_0^\prime$ is the wavevector of an
electron in the substrate probe, of the same energy $E$.

In the present work, molecular orbitals
are evaluated using extended H\"{u}ckel theory\cite{yaehmop}
and therefore
require a non-orthogonal basis set within the molecule. It
has been shown that a simple change of Hilbert space can
redefine the problem in terms of a system with an orthogonal
basis\cite{Emberly98}. This is achieved by transforming the Hamiltonian of
the system into a new energy-dependent Hamiltonian $H^E$:
\begin{equation}
H^E = H - E(S - I)
\label{Horthog}
\end{equation}
where $H$ is the original Hamiltonian matrix, $S$ is the
overlap matrix, and $I$ is identity. In the model presented
here, we assume orthogonality between the orbitals of the
probe leads, although by using Eq.(\ref{Horthog}) the
model could easily be extended to systems where
these orbitals of the probe are non-orthogonal. 

By using the Lippmann-Schwinger approach, we are free to
choose convenient boundaries for the central scattering
region, not necessarily restricted to the actual molecule.
In order to model the coupling between the probes and
molecule in a realistic way, we consider the probe atoms
that are closest to the molecule to be part of
an {\it extended molecule} (see Fig. \ref{fig3}), i.e.,
we treat them as if they were parts of the
molecule. Their orbitals $|a\rangle$ and $|b\rangle$
are assumed to be orthogonal to the lead orbitals
$|-1\rangle$ and $|1\rangle$ on the lead sites adjacent
to them.

Then, we have
\begin{align}
W_{-1,j}&=W_{j,-1}=\langle -1|H|a\rangle\langle a|\phi_j\rangle
=\beta c_{a,j}\nonumber \\
W_{j,1}&=W_{1,j}=\langle \phi_j|b\rangle\langle b|H|1\rangle
=\beta c _{b,j}.
\end{align}
In order to calculate the electric current passing through an
STM/molecule/substrate system, the transmission probability
of an electron, $T(E)$, is integrated through the energy
range inside the Fermi energy window between
the two probes that is created when a bias voltage is applied.
To obtain a theoretical STM current map, this
electric current calculation is performed for many
different positions of the tip probe,
while the substrate probe remains stationary. The simplicity
of this model allows a complete current map to be
generated in a reasonable amount of time. By comparing current
maps that are
generated for different substrate probe configurations,
we are able to develop an intuitive understanding of
the important role substrates may play in STM experiments
on single molecules. 

In the remainder of this paper, we will consider, as an example,
a molecule of current experimental interest\cite{Qiu03}, 
Zn(II)-etioporphyrin I. For simplicity, we model
the probes as consisting of Cu s-orbitals,
and compare various simulated constant-height STM current maps
of the molecule obtained using different substrate probe
locations, corresponding to different possible locations of
dominant molecule-substrate coupling.
We will demonstrate how the properties of an STM current image
may display a remarkable qualitative dependence on the location
of this molecule-substrate coupling.
\newpage
\section{Model Results}

We present results for the single-molecule system
of Zn(II)-etioporphyrin I (ZnEtioI) (see Fig. \ref{fig4}),
coupled to model tip and substrate probes that we
represent for simplicity by Cu s-orbitals.
Density functional theory
was used in obtaining the geometrical structure of
ZnEtioI\cite{Gaussian98}.
The molecule is mainly planar, but contains
4 out-of-plane ethyl groups. 

The electronic structure of the molecule was computed using
the extended H\"{u}ckel model\cite{yaehmop}.
In this model, the energy
of the highest occupied molecular orbital (HOMO) was
found to be -11.5 eV, and the energy of the lowest
unoccupied molecular orbital (LUMO) was found to be
-10.0 eV. The Fermi level of a metallic probe in contact
with a molecule at zero applied bias is usually located
between molecular HOMO and LUMO levels. However,
establishing the precise position of the Fermi energy of the
probes relative to the HOMO and LUMO is in general a
difficult problem in molecular electronics, with different
theoretical approaches yielding differing
results\cite{2Emberly98,DiVentra,Damle}.
Therefore, within this illustrative model, we consider
two possible zero-bias Fermi energy positions for the probes: 
In the {\it LUMO-energy transmission} subsection (\ref{LUMO}),
the Fermi energy
is taken to be -10.4 eV\cite{cluster}.
Thus, at $V_{bias}=1.0$ V, the
Fermi energy window will include the LUMO but not the HOMO.
In the {\it HOMO-energy transmission} subsection (\ref{HOMO}),
the Fermi energy
is taken to be -11.4 eV. In this case, at $V_{bias}=1.0$ V,
the Fermi energy window will include the HOMO but not the
LUMO.\cite{biasdirection}

\subsection{LUMO-energy transmission}
\label{LUMO}
We first consider the case of transmission through the
molecule at LUMO energies. For this, we set
$V_{bias}=1.0$ V, with $E_F=-10.4$ eV at zero bias.
The substrate probe is now positioned to
simulate various possible locations of dominant
molecule-substrate couplng. Four different positions
for the substrate probe are analyzed, as shown by the blue
circles in Fig. \ref{fig4}:
directly below one of the outer ethyl groups of the molecule (A),
below an inner carbon atom of the molecule (B), below a nitrogen
atom (C), and below the zinc center of the molecule (D).
The orbital representing the substrate probe, in each
case, is centered $2.5$\AA\space below the nearest atom in the
molecule. Constant-height STM current images for these
substrate probe positions are simulated by moving the tip
probe across the molecule in steps of $0.25$\AA, calculating
the electric current at each step, thus creating a
$16$\AA $\times 16$\AA\space electric current image
(transmission pattern). The
tip probe in all cases is located $2.5$\AA\space above the plane of
the molecule. 

Fig. \ref{fig5}(a,b,c,d) shows the simulated current images obtained
in each case, the blue circle indicating the position of the
substrate probe. Each image has unique features not seen in the
other images, that arise due to differences in the details of the
molecule-substrate coupling. In Fig. \ref{fig5}(a), with the substrate
probe positioned below an outer ethyl group as shown in Fig. \ref{fig4}
(position A),
a delocalized transmission pattern is
obtained. A localized region of enhanced transmission exists
where the tip probe is directly above the same ethyl group that is
coupled to the substrate probe. In Fig. \ref{fig5}(b), a somewhat similar
transmission lobe pattern is obtained, with the substrate probe
positioned below an inner carbon atom (see Fig. \ref{fig4}, position B).
In this configuration, however, the transmission pattern has two-fold
symmetry and there is no apparent localized region of enhanced
transmission. Furthermore, the lobes of high transmission in
Fig. \ref{fig5}(b) are 1-2 orders of magnitude stronger than the corresponding
lobes in Fig. \ref{fig5}(a), as will be discussed below. In the case when
the substrate probe is directly below a nitrogen atom (see Fig. \ref{fig4},
position C), a distinct transmission pattern is obtained, shown in
Fig. \ref{fig5}(c). The lobe with the highest transmission in this figure
is 1-2 orders of magnitude weaker than lobes seen in Fig. \ref{fig5}(b). 
In this case, a localized region of enhanced transmission exists
where the tip probe is above the same nitrogen atom that
is coupled to the substrate probe.
Fig. \ref{fig5}(d) shows a
very different transmission
pattern. In this case, the substrate probe is positioned directly
below the center zinc atom of the molecule (Fig. \ref{fig4}, position D),
and transmission is
found to occur primarily when the tip probe is above the center of
the molecule. 

In order to help understand the differences between these images, the
characteristics of the LUMO were investigated. The LUMO is
a degenerate $\pi$-like orbital with two-fold symmetry.
Analyzing the LUMO as a linear combination of atomic orbitals,
we find that contributions to the LUMO come primarily from atomic
orbitals in the core porphyrin structure, with low contributions
from the ethyl groups and the central zinc atom. Particularly high
contributions come from two of the four inner corner carbon
atoms (the atom above substrate probe B and the corresponding atom
180 degrees away, in Fig. \ref{fig4}, or the
equivalent atoms under rotation of 90 degrees for the other
degenerate LUMO orbital).
Therefore, in the case of Fig. \ref{fig5}(b), there is a strong coupling between
the substrate probe and one of the two degenerate LUMOs of the molecule,
whereas in the case of Fig. \ref{fig5}(a), with the substrate probe below the ethyl
group, there is only a weak substrate-LUMO coupling. This explains why
the transmission pattern of Fig. \ref{fig5}(b) is much stronger than Fig. \ref{fig5}(a).
Regarding the similar appearance of the transmission patterns in the two
cases, we expect LUMO-mediated transmission to occur, in both cases, when
the tip probe has significant
coupling to the LUMO. The delocalized transmission patterns of Fig. \ref{fig5}(a)
and Fig. \ref{fig5}(b) in fact correspond well to areas of high atomic orbital
contributions to the LUMO, with the low-transmission nodes occuring in
regions of the molecule where the amplitude of the LUMO is close to zero.

The differences between the transmission
patterns may be better understood by studying T(E) for appropriate tip
probe positions in each case. Fig. \ref{fig5}(e,f,g,h) shows T(E) for the 
corresponding placement of the tip probe as labelled by red dots in
Fig. \ref{fig5}(a,b,c,d). In Fig. \ref{fig5}(e), T(E) is shown for the
localized region of enhanced transmission in Fig. \ref{fig5}(a).
There is a transmission resonance associated with the LUMO (at -10 eV),
together with an antiresonance that occurs at a slightly lower energy.
The antiresonance, along with antiresonances seen in subsequent
figures (with the exception of the antiresonance in Fig. \ref{fig5}(f)),
arises due to interference between electron propagation
through a weakly coupled orbital (in this case the LUMO) and propagation
through other orbitals of different energies. This can be seen
mathematically through Eq.(\ref{transmit1}) and Eq.(\ref{ABCD}). Transmission
drops to 0 when $A = 0$. This occurs when all the terms
$W_{1,j}(\frac{1}{E-\epsilon_j})W_{j,-1}$ for the different orbitals sum
to 0. If an orbital is weakly coupled to the probes, its contribution to $A$
is small unless the electron energy is close to the energy of the orbital.  
When the electron energy does approach this orbital energy, the
contribution to $A$ will increase and, if its sign is opposite,
cancel the other orbitals' contributions. Thus, these types of
antiresonances are always seen on only one side of a transmission peak
of a weakly coupled orbital. Returning to Fig. \ref{fig5}(e), we see that,
although transmission via the LUMO contributes some of the electric current,
a significant contribution comes from the background. We find this background
transmission to be composed primarily of the high energy
transmission tails of molecular orbitals localized on the ethyl groups.
When the tip probe is coupled to the same ethyl group as the substrate
probe, transmission via these ethyl-composed molecular orbitals
is strong and has a significant tail extending to the relevant
range of energies near the LUMO. Fig. \ref{fig5}(f) shows T(E) for the same
tip probe position as Fig. \ref{fig5}(e), but with the substrate probe positioned
below an inner carbon atom, as in Fig. \ref{fig5}(b). Since the substrate probe
is not significantly coupled to the ethyl group, the ethyl-based
transmission background is negligible, and the region of locally enhanced
transmission seen in Fig. \ref{fig5}(a) is not seen
in Fig. \ref{fig5}(b). It should also
be noted that the transmission peak in Fig. \ref{fig5}(f) is wider than in
Fig. \ref{fig5}(e), due to hybridization of the LUMO with the strongly coupled
substrate probe. The antiresonance seen at the center of the peak
is due to the degeneracy of the LUMO. In this case, one of the
LUMO orbitals is strongly coupled to the substrate probe,
with the other being only weakly coupled. The weakly coupled orbital
causes electron backscattering to occur, resulting in an antiresonance
at the LUMO energy.
In Fig. \ref{fig5}(g), the substrate probe is directly
below a nitrogen atom and the tip probe directly above. In this case, the
transmission peak corresponding to the LUMO is very narrow, and current flow
comes primarily from background transmission. This background transmission
corresponds mainly to the high energy transmission tails of molecular orbitals
that have strong contributions from the nitrogen atoms. The transmission pattern
seen in Fig. \ref{fig5}(c) is the result of contributions from these various
low-energy orbitals, and from the HOMO$-$1 and HOMO$-$2, which will be analyzed in
greater detail in subsection \ref{HOMO}. 
Transmission through the LUMO is quenched because the substrate probe
is coupled to a region of the molecule where the amplitude of the LUMO is close
to zero. Thus, the overall transmission pattern is weak compared to
Fig. \ref{fig5}(b). 
In Fig. \ref{fig5}(h), the substrate probe is directly below
the center of the molecule and the tip probe directly above. For this
case, the transmission curve contains no LUMO-related transmission
peak, since the LUMO is an antisymmetric orbital and has a node at
the center of the molecule. Instead, we see a transmission background
that rises
smoothly with energy. This transmission corresponds to the tail of a
higher-energy $\pi$-like orbital composed primarily of zinc, with
additional, less-significant contributions from other atoms. 
The transmission pattern of Fig. \ref{fig5}(d), plotted on a log scale,
is shown in Fig. \ref{fig6}, and reveals
additional structure of this orbital. Transmission through this orbital
has delocalized features not evident in Fig. \ref{fig5}(d), such as nodes of
low transmission when the tip probe is above a nitrogen atom, as well
as regions of higher transmission when the tip probe is above the outer
sections of the molecule.
In Fig. \ref{fig5}(h), the probes are both coupled strongly to this orbital,
so the orbital hybridizes with
the probes and creates a transmission peak with a very long tail. Compared
to this tail, transmission via the LUMO (which has very low zinc content)
is negligible.

\subsection{HOMO-energy transmission}
\label{HOMO}
Next, we consider electron transmission at energies close to the HOMO.
For the purposes of analyzing HOMO-mediated transmission,
we consider the probes to have a zero-bias Fermi energy of $-11.4$ eV, which
is closer to the HOMO than the LUMO. We again set $V_{bias}=1.0$ V, and
consider the same four cases of substrate probe position as for
transmission at LUMO energies.

The HOMO of zinc-etioporphyrin is a non-degenerate $\pi$-like orbital
with 4-fold symmetry and an energy of -11.5 eV. The primary atomic
contributions to this orbital are from carbon atoms in the 4
pyrole rings, with weak contributions from the ethyl groups
and negligible contributions from all of the other inner atoms. In the
energy window we are considering, there exists another $\pi$-like
orbital (HOMO$-$1), also 4-fold symmetric and with an energy of -11.8 eV.
Unlike the HOMO, this orbital has large contributions from the inner corner
carbon atoms (see Fig. \ref{fig4}, above position B, and symmetric equivalents).
It also has significant contributions from the
nitrogen atoms, as well as non-negligible contributions
from the zinc center and the 4 ethyl groups. In this energy range, there
is also a $\sigma$-like orbital (HOMO$-$2) at an energy of -11.9 eV,
with strong contributions from the nitrogen atoms.

Transmission patterns for this energy range are shown in Fig. \ref{fig7}(a,b,c,d),
corresponding to the same substrate probe positions as in Fig. \ref{fig5}(a,b,c,d).
In the case where the substrate probe is directly below an ethyl group
(Fig. \ref{fig7}(a)), a complex transmission pattern is obtained. In particular,
low-transmission nodes exist every 45 degrees. To understand the source
of these nodes, T(E) is shown (see Fig. \ref{fig7}(e)) for two different tip probe
positions that are very close to each other, one being directly on a node
(the red dot in Fig. \ref{fig7}(a))
and the other a small distance away but in a region of higher transmission
(the black dot). Note that T(E) is shown, in this case only, in the
narrower energy range of -11.9 eV to -11.4 eV. (No transmission peaks
are present in the energy range from -11.4 eV to -10.9 eV.)
We see that transmission through the HOMO is extremely quenched (the
transmission peak narrows) when the
tip probe is above the node, but transmission through the HOMO$-$1 is
relatively unaffected. (The very narrow -11.88 eV transmission peak
corresponding to the $\sigma$-like HOMO$-$2 orbital
has a negligible effect on overall current flow.)
This quenching of transmission through the HOMO
occurs because the tip probe is closest to a region of the molecule where
the HOMO's amplitude is nearly zero. These regions occur every 45 degrees,
as shown by the nodes.
The other (curved) low-transmission nodes that are seen in Fig. \ref{fig7}(a)
are caused by the
HOMO$-1$, as will become clear through analysis of Fig. \ref{fig7}(b). Since both
the HOMO and HOMO$-$1  are coupled
non-negligibly to the substrate probe in Fig. \ref{fig7}(a),
we see a transmission pattern that
is affected by both of these orbitals.
In the case (Fig. \ref{fig7}(b)) when the substrate probe is below an inner corner
carbon atom (Fig. \ref{fig4}, position B),
a transmission pattern that is significantly different
from Fig. \ref{fig7}(a) is obtained. The low-transmission nodes every 45 degrees are not
seen, and there are strong transmission peaks when the tip probe is above
one of the 4 inner corner carbon atoms. In Fig. \ref{fig7}(f), T(E) is shown for
the case when the tip probe and substrate probe are directly above and below
the same corner carbon atom. The HOMO$-$1 is clearly the dominant pathway
for transmission through the molecule, with the HOMO and HOMO$-$2 producing only
narrow additional transmission peaks. This is understandable, since the corner
carbon atom which is closest to both the tip and substrate probes has a negligible
contribution to the HOMO, but a large contribution to the HOMO$-1$. Hence,
the transmission pattern seen in Fig. \ref{fig7}(b) is primarily due to
(HOMO$-1$)-mediated transmission through the molecule. The curved low
transmission nodes correspond to regions of the molecule where the
amplitude of the HOMO$-$1 is close to 0. Similar curved low-transmission
nodes are also seen in Fig. \ref{fig7}(a), illustrating that the
HOMO$-$1 is also the source of these nodes.
In the case when the substrate probe is below a nitrogen atom, another unique
transmission pattern is obtained. In Fig. \ref{fig7}(g), T(E) is shown for
the case when the tip probe and substrate probe are above and below the same
nitrogen atom. Two transmission peaks of similar strength are seen, corresponding
to the HOMO$-$1 and HOMO$-$2, as well as a very weak peak corresponding to the
HOMO. This is understandable, since both the HOMO$-$1 and HOMO$-$2 have considerable
nitrogen contributions, and the HOMO does not.
Hence, the transmission pattern seen in Fig. \ref{fig7}(c) is due to both
the HOMO$-$1 and HOMO$-$2, resulting in a unique transmission pattern.
Lastly, when the substrate probe is below the center of the molecule
(Fig. \ref{fig7}(d)), a transmission pattern looking quite similar to Fig. \ref{fig7}(b)
is obtained. Unlike in the case of LUMO energies, the transmission pattern
for HOMO energies is not dominated by transmission through the low-energy
tail of a zinc-dominated orbital. Rather, transmission appears to be
mediated mainly by the HOMO$-$1 orbital. This is because the HOMO$-$1,
unlike the HOMO or LUMO, has non-negliglble contributions from the center
zinc atom, that is strongly coupled to the substrate probe in this case.
In Fig. \ref{fig7}(h), T(E) is shown for the case of the tip probe and substrate
probe being directly above and below the center of the molecule. We see
a main transmission peak corresponding to the HOMO$-$1, as well as a background
due to the tail of the higher-energy zinc-dominated orbital. This results
in stronger transmission when the tip is above the center of the molecule
than if only the HOMO$-$1 is strongly coupled to the substrate probe,
as occurs in Fig. \ref{fig7}(b). 

All of the unique features seen in each of these four cases, for both
HOMO and LUMO energy ranges, directly
arise from differences in the details of the
molecule-substrate coupling in each case.
While an individual substrate probe positioned below the molecule is an
incomplete representation for the molecule-substrate interaction,
this representation illustrates the importance of understanding the
detailed nature of the molecule-substrate interaction when analyzing
and modeling STM topographs of single molecules on substrates.
Nevertheless, specific experimental results can indeed be
shown to be consistent with results of the model presented in
this article, as will be discussed next.

\subsection{Comparison with Experiment}
\label{Comparison with Experiment}
STM transmission patterns for the system of
Zn(II)-etioporphyrin I adsorbed on
inhomogeneous alumina covering a NiAl(110) substrate
have recently been obtained experimentally\cite{Qiu03}.
These experimental results generally show
four lobes above the etioporphyrin molecule,
where placement of the STM tip results in high transmission. 
Experimentally, the relative transmission through each
of the lobes is found to depend strongly on which individual
molecule is being probed, due to the complex
nature of the alumina-NiAl(110) substrate.
Often, one or two lobes are found to have much higher
transmission than the rest. These asymmetries were originally
attributed to conformational differences between molecules.
However, a further investigation of conformational differences
only identified different molecular conformations that
produce {\it two-fold symmetric} patterns\cite{Qiu04b}. Thus,
a different explanation is needed for the images of lower
symmetry seen on the alumina. 

An alternate explanation for the various different STM
images obtained for individual molecules will now be presented.
In the experiments, the molecules were likely more
strongly coupled to the substrate than to the STM tip, since the
molecules were adsorbed on the substrate,
and the experiments were performed at a
relatively low tunneling current of 0.1 nA. The STM images were
obtained at positive substrate bias, therefore we may infer 
that the lobes represent regions of strong transmission around
{\it LUMO energies}. The experimental results are
consistent with the two-probe model results
for the situation shown in Fig. \ref{fig5}(a) (at LUMO energies,
with the substrate probe
placed below one of the out-of-plane ethyl groups of the molecule),
as will be explained below.
To more realistically
model what one might see in an STM experiment
with finite lateral resolution, the resolution of
Fig. \ref{fig5}(a) should be reduced: Fig. \ref{fig8} shows
the same transmission pattern as Fig. \ref{fig5}(a), but in convolution
with a gaussian weighting function of width $6$\AA. We see that 
two distinct high transmission lobes emerge, one much stronger than
the other, about $11$\AA\space apart.
Experimentally, the most common image seen by
Qiu et al. (Fig. 2B in their article\cite{Qiu03}) is, after an
appropriate rotation, remarkably similar to Fig. \ref{fig8},
also containing two dominant asymmetric lobes, 
located $11$\AA\space apart.

The other less-common STM images observed experimentally can
also be explained qualitatively with our model.
In an experimental situation, the underlying metal substrate may be
coupled to {\it all four} ethyl groups at significantly differing
strengths depending on the detailed local arrangement and strengths of the
most conductive spots of the alumina film (discussed in Section
\ref{Introduction}) in the vicinity of the molecule. The result would
resemble a superposition of Fig.
\ref{fig8} and current maps derived from Fig. \ref{fig8} by rotation
through 90, 180, and 270 degrees, with weights depending on the relative
strength of the coupling of the substrate to each of the ethyl
groups.\cite{incoherent} In this analysis, other substrate
probe positions that are the same distance from the plane of the molecule
(about $4$\AA)
but not below an ethyl group have also been considered.
It was found that other
substrate probe positions yielded much weaker current flow through the
molecule.
Thus, these positions can be neglected in a first approximation, and
current flow can be assumed to be dominated by pathways through
the four substrate probe positions below the ethyl groups.
All of the different transmission
pattern results obtained experimentally can be reproduced in this way
reasonably well, given the simplicity of the model and the
fact that the model results are for constant-height calculations whereas
experimentally, constant-current STM images are obtained. 

One final consideration is that in an experimental situation, the out-of-plane
ethyl groups of the Zn-etioporphyrin molecule may possibly point
{\it away} from the substrate, contrary to what has been assumed above.
Thus, we now consider this case. Fig. \ref{fig9} shows transmission
patterns that correspond to the four substrate probe positions
shown in Fig. \ref{fig4}, assuming the ethyl groups point {\it away}
from the substrate probe. The substrate probe is
positioned $2.5$\AA\space below
the plane of the molecule, and the tip probe scans the molecule
at a constant height of $4$\AA\space above the plane. 
We see that in the case of Fig. \ref{fig9}(a), two asymmetric lobes
corresponding to the out-of-plane ethyl groups dominate the image,
one about double the strength of the other. In
Fig. \ref{fig9}(b,c), two symmetric ethyl-based lobes dominate the
images, with strengths similar to the strength of the weaker lobe
of Fig. \ref{fig9}(a). In Fig. \ref{fig9}(d), however, current
flows primarily through the center of the molecule, again with a
strength similar to that of the weaker lobe of \ref{fig9}(a).
Thus, we see that most substrate probe positions (other than
below the center of the molecule) produce current patterns
with high-transmission lobes corresponding to the locations of the
ethyl groups, with the strongest current pattern, obtained when
the substrate probe is below an ethyl group, producing
asymmetric lobes. Therefore, with the assumption that the ethyl groups
of the molecule point away from the substrate, the different
transmission pattern results obtained experimentally, showing four
asymmetric lobes, can clearly still be reproduced within our model.

\newpage
\section{Conclusions}

We have explored theoretically a model of scanning tunneling
microscopy in which a molecule is contacted with
two {\it local} probes, one representing the STM tip and the
other the substrate. This is the simplest model of STM of
large molecules separated from conducting substrates by
thin insulating films where the dominant
conducting pathway through the insulating film is localized
to a region smaller than the molecule.

We have applied this model to Zn(II)-etioporphyrin I molecules
on a thin insulating alumina layer.
In recent experiments on this system,
very different topographic maps were obtained for 
molecules at different locations on the substrate.
We have shown that differences in the
details of the effective molecule-substrate coupling due to the
non-uniform transmission of electrons through the alumina
can account for the differences in topographic maps of these molecules.
Our model results suggest that the out-of-plane ethyl groups
of the molecule may be the location of dominant
molecule-probe coupling. 

Our theory also suggests that further experiments in which the
molecules are on a thin alumina film over an NiAl(111) substrate
(complementing the work in Ref.
\onlinecite{Qiu03} with the NiAl(110) substrate) would be of interest:
Unlike thin alumina films on NiAl(110) substrates,\cite{NiAL110} thin
alumina films on NiAl(111) substrates have $periodic$ arrays of
spots at which electron transmission through the alumina is
enhanced.\cite{NiAL111} Thus for Zn(II)-etioporphyrin I molecules on
alumina/NiAl(111) it may be possible to observe simultaneously both the
periodic array of spots where transmission through the alumina is
enhanced and the STM images of molecules on the surface and to study
experimentally the interplay between the two in a controlled way. 

Studying the scanning tunneling microscopy
of molecules using a framework of
two local probes opens a new avenue
for future theoretical and experimental research,
and we hope that it will help to achieve a
greater understanding of molecular electronic systems.

\section*{Acknowledgments}

This research was supported by NSERC and the
Canadian Institute for Advanced Research.

%FIGURES

\begin{figure}[!t]
\caption{Illustrative diagram for an STM/molecule/substrate experiment,
showing a possible pathway for electron transmission
when the molecule is weakly bound to the substrate due to the
presence of a complex insulating layer that transmits electrons non-uniformly. A region of
dominant molecule-substrate coupling causes electron transmission to
occur primarily through a single pathway.}
\label{fig1}
\end{figure}

\begin{figure}[!t]
\caption{Illustration of a two-probe theory of scanning tunneling
microscopy. Tip and substrate are both considered to be
local probes coupled to the molecule.}
\label{fig2}
\end{figure}

\begin{figure}[!t]
\caption{A schematic diagram of the model STM/molecule/substrate
system. The tip and substrate probes are semi-infinite. Nearest
neighbour atoms to the molecule (with atomic orbitals labelled
$|a\rangle$ and $|b\rangle$) are considered to be part of the
{\it extended molecule}, which is represented by the dashed
rectangle.}
\label{fig3}
\end{figure}

\begin{figure}[!t]
\caption{(color online). The Zn(II)-etioporphyrin I molecule. Carbon atoms
are red, nitrogen atoms are green, the zinc atom is yellow,
and hydrogen is white. The four blue circles
(labelled A, B, C, and D) denote four possible positions 
for the substrate probe below the molecule (into the page), 
that are considered in this article. In each case, the closest
atom in the substrate probe (atomic orbital $|b\rangle$ in
Fig. \ref{fig3}) is $2.5$\AA\space below the nearest atom
of the molecule.}
\label{fig4}
\end{figure}

\begin{figure}[!t]
\caption{(color online). Transmission at LUMO
energies. a,b,c,d)$16$\AA $\times 16$\AA\space constant-height
transmission patterns, for 4 different substrate probe positions.
Darker regions represent tip probe positions that give higher current
flow through the molecule. The blue
circles represent the position of the substrate probe below the
molecule in each case, the closest atom of the probe being
$2.5$\AA\space below the nearest atom of the molecule.
These positions correspond to the blue
circles in Fig. \ref{fig4}: a)circle A, b)circle B, c)circle C, d)circle D.
The red dots represent tip probe positions for the
corresponding T(E) curves shown in e,f,g,h respectively.
e) Transmission vs. energy
for tip and substrate probes directly above and below an
outer ethyl group. f) T(E) for the tip probe above the same
ethyl group but the substrate probe below an inner corner carbon atom.
g) T(E) for tip and substrate probes above and below a nitrogen atom.
h) T(E) for tip and substrate probes above and below the central
zinc atom. In all cases, the tip probe is $2.5$\AA\space above
the plane of the molecule.}
\label{fig5}
\end{figure}

\begin{figure}[!t]
\caption{The same $16$\AA $\times 16$\AA\space transmission
pattern shown in Fig. \ref{fig5}d, with
transmission plotted
on a Log scale. Additional delocalized features can be seen.
}
\label{fig6}
\end{figure}

\begin{figure}[!t]
\caption{(color online) Transmission at HOMO
energies. a,b,c,d)$16$\AA $\times 16$\AA\space constant-height
transmission patterns, for 4 different substrate probe positions.
As in Fig. \ref{fig5}, the blue circles represent the position
of the substrate probe below the molecule. Again, they correspond to
the blue circles in Fig. \ref{fig4}: a)circle A, b)circle B,
c)circle C, d)circle D. The red dots represent tip probe positions for the
corresponding T(E) curves shown in e,f,g,h respectively. An additional
black dot
just below the red dot in (a) represents a different tip probe
position, yielding a second T(E) curve in e). e) T(E) for the
substrate probe directly below an outer ethyl group, and the tip
probe on a low transmission node [red curve, red dot in (a)],
or close to this node [black curve, black dot in (a)]. The narrow
transmission peak near -11.9 eV exists for both curves (the black
curve is under the red curve). Note
that the energy scale is different than for f,g,h. f) T(E) for the tip
and substrate probes above and below an inner corner carbon atom.
g) T(E) for tip and substrate probes above and below a nitrogen atom.
h) T(E) for tip and substrate probes above and below the
central zinc atom. In all cases, each probe
is $2.5$\AA\space away from the
nearest atom in the molecule.}
\label{fig7}
\end{figure}

\begin{figure}[!t]
\caption{The $16$\AA $\times 16$\AA\space transmission
pattern shown in Fig. \ref{fig5}a, in convolution
with a gaussian weighting function of width $6$\AA. This is
done in order to more realistically simulate what one might
expect to see in a real STM experiment. Two distinct
asymmetric lobes are visible, and the calculated pattern is similar to the
most common STM image observed experimentally by Qiu $et al.$\cite{Qiu03}
which is shown in Fig. 2B of their paper.}
\label{fig8}
\end{figure}

\begin{figure}[!t]
\caption{Transmission at LUMO energies, assuming ethyl groups point
{\it away} from the substrate
probe. a,b,c,d)$16$\AA $\times 16$\AA\space transmission
patterns for the 4 different substrate probe
positions shown in Fig. 4: a)circle A, b)circle B, c)circle C, d)circle D.
The substrate probe is in all cases $2.5$\AA\space below 
the plane of the
molecule, and the tip probe is $4.0$\AA\space above
the plane of the molecule.}
\label{fig9}
\end{figure}


\begin{references}
\bibitem{drop}In this work we are considering thin, mainly planar
molecules oriented parallel to the substrate. In this geometry most of
the potential drop occurs in the insulating layer between the metal
substrate and the molecule, and between the molecule and the STM tip.
Thus to a first approximation the potential drop within the molecule
itself can be neglected.
\bibitem{cluster} This value is predicted by an extended H\"{u}ckel
cluster calculation for a Cu probe.
\bibitem{biasdirection} As is discussed in Section \ref{Comparison with
Experiment}, in the relevant experiments,\cite{Qiu03} the
molecular LUMO (HOMO) state is probed when the tip is biased negatively
(positively) relative to the substrate.
\bibitem{incoherent}Here we have assumed for simplicity 
that processes injecting electrons into different ethyl groups
are mutually incoherent.
\end{references}
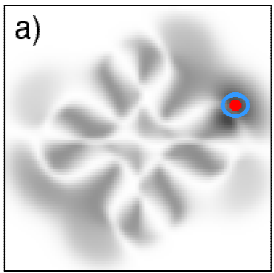
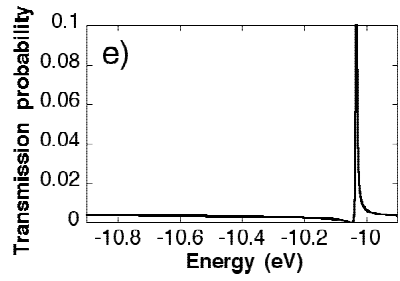
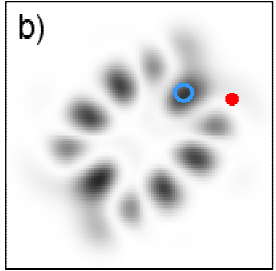
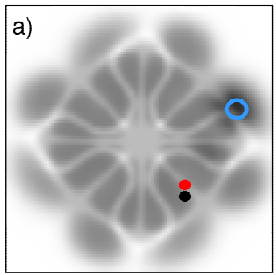
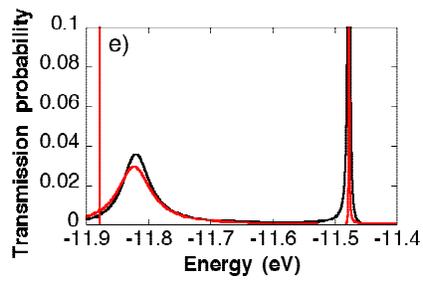
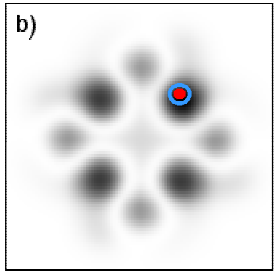
\end{document}